\documentclass{article}

\usepackage{arxiv}

\usepackage[utf8]{inputenc} 
\usepackage[T1]{fontenc}    
\usepackage{lmodern}
\usepackage{hyperref}       
\usepackage{url}            
\usepackage{booktabs}       
\usepackage{amsfonts}       
\usepackage{nicefrac}       
\usepackage{microtype}      
\usepackage{lipsum}
\usepackage{fancyhdr}       
\usepackage{graphicx}       
\graphicspath{{media/}}     
\usepackage{amsmath}
\usepackage{booktabs}
\usepackage{multirow}
\usepackage{xcolor}

\usepackage{listings}
\usepackage{color}
\usepackage{dsfont}

\definecolor{dkgreen}{rgb}{0,0.6,0}
\definecolor{gray}{rgb}{0.5,0.5,0.5}
\definecolor{mauve}{rgb}{0.58,0,0.82}

\title{Voice Communication Analysis in esports}

\author{
 Aymeric Vinot \\
  Independent researcher\\
  Data Scientist/Analyst\\
  $\mathbb{X}$ : \texttt{@Warock42}\\
  \texttt{aymericvinot38@gmail.com} \\
   \And
 Nicolas Perez \\
  GiantX esport \\
  Assistant coach \\
  $\mathbb{X}$ : \texttt{@PerezNicolasLol}\\
  \texttt{mail.perez.nicolas@gmail.com} \\
}

\begin{document}
\maketitle

\begin{abstract}
In most team-based esports, voice communications are prominent in the team efficiency and synergy. In fact it has been observed that not only the skill aspect of the team but also the team effective voice communication comes into play when trying to have good performance in official matches. With the recent emergence of LLM (Large Language Models) tools regarding NLP (Natural Language Processing) \cite{vaswani2017attention}, we decided to try applying them in order to have a better understanding on how to improve the effectiveness of the voice communications. In this paper the study has been made through the prism of League of Legends esport. However the main concepts and ideas can be  easily applicable in any other team related esports.
\end{abstract}

\keywords{Voice analysis \and esport \and NLP}

\section{Introduction}
In most team-based esports, voice communications are prominent in the team efficiency and synergy. In fact it has been observed that not only the skill aspect of the team but also the team effective voice communication comes into play when trying to have good performance in official matches. With the recent emergence of LLM (Large Language Models) tools regarding NLP (Natural Language Processing) \cite{vaswani2017attention}, we decided to try applying them in order to have a better understanding on how to improve the effectiveness of the voice communications. In this paper the study has been made through the prism of League of Legends esport. However the main concepts and ideas can be  easily applicable in any other team related esports.\\
	Today, this aspect of voice analysis in esport is overlooked due to the lack of tools an shared knowledge that could help solve this problem. Moreover this subject is a very interesting study case. In fact we need to employ few-shot techniques to perform our study on esport voice communication as very few to none datasets are publicly available. This is why we conducted this analysis in order to find some starting tracks in this field. The main objectives of such voice communication analysis are to build metrics to determine how effective players are communicating during the game. In fact building such evaluation tools could be very relevant to correlate with in-game performance metrics, thus trying to pin-point the positive and/or negative potential impacts of communication quality in the overall game performance. After some surveys with coaches in professional teams we ended up with two main issues regarding the communication effectiveness :
\begin{itemize}
	\item \textbf{Duplicate communications}: Sometimes players are communicating the same idea several times in a short period of time. Hence blurring the conversation and reducing the effectiveness of the communication.
	\item \textbf{Parasite communication}: Sometimes players communicate ideas that are unclear/not relevant in the context of the game.
\end{itemize}
This article aims to propose possible solutions for these two problems that will be respectively treated in section \ref{sec:DuplicateCommunications} and section \ref{sec:ParasiteCommunications}

\section{Audio processing pipeline}
\begin{figure}[h!]
	\centering
	\includegraphics[width=0.85\textwidth]{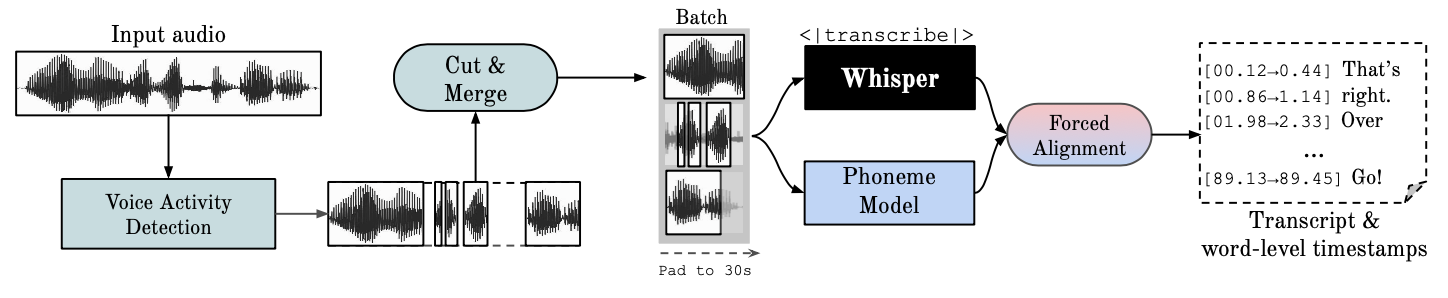}
	\caption{Pipeline of audio transcription from \cite{bain2022whisperx}}
	\label{fig:whisperx}
\end{figure}

Just before diving into the details of our solutions, we wanted to share with you how we treated such audio files. For this we used the work of Bain et. al.\cite{bain2022whisperx} to transcribe the audio into text. The method used in \cite{bain2022whisperx} can be seen at Figure \ref{fig:whisperx}

Thus, with the help of \cite{bain2022whisperx} and some adjustments, the following has been done:\\
\begin{enumerate}
	\item First we transcribe the audio file using Whisper \cite{openai2022whisper} from OpenAI.
	\item Then we perform speaker diarization with the help of PixIT model \cite{Kalda24}   implemented by pyannote.audio \cite{Bredin23}.
	\item Finally we perform forced-text alignment to align words spoken by each player with its timestamp.
\end{enumerate}

However it is possible to use direclty other pieces of software (discord bots and so on) to skip the uncertainty of the speaker-diarization step. As we observed that some pieces of speech were wrongly attributed to a given speaker.

\section{Duplicate communications}
In this section we will present an approach of repetitive/duplicate communication detection based on lexical similarity. By using this approach, we move closer to the process of retrieving relevant pieces of information in open-domain question answering \cite{Chen2017odqa} by checking how similar two pieces of text are by semantic/lexical meaning. One of the advantage of using sentence similarity is being able to have easily interpretable and non-chaotic results.
\label{sec:DuplicateCommunications}

\begin{figure}[h!]
	\centering
	\includegraphics[width=0.85\textwidth]{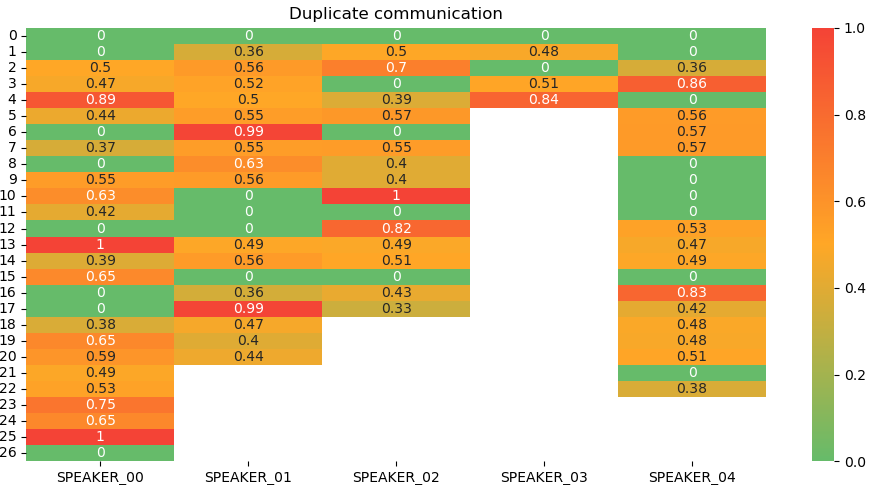}
	\caption{Duplicate communication scores on each sentence for each speaker. Each sentence is assigned a score ranging from 0 to 1 telling how much this said sentence is repetitive in terms of semantic similarity compared to the previous sentences spoken in a time frame of $W = 15s$}
	\label{fig:duplicate-communication}
\end{figure}

\subsection{Explanation and solution of the problem}
\label{duplicate-com-explanation}

\subsubsection{Formal explanation and solution}
Traditionally, when comparing the similarity between two pieces of text to leverage lexical similarities, we often use TF-IDF or BM25 weighting \cite{robertson2009bm25}. However these approaches based on near-exact matches between keywords between two pieces of text suffer from the lexical gap and do not generalize well \cite{berger}. By contrast, approaches based on neural networks allow learning beyond lexical similarities, resulting in way better performances for our task. We will refer to this method by "semantic similarity". This metric will be used on each sentences and compare it to every other previous sentences spoken by the given player in the last $W$ seconds. For example, consider the following snippets of conversation :

\begin{verbatim}
...
010 - [082.458:084.699] SPEAKER_00 I mean in 3:00, we have to watch out for Zyra though.
011 - [091.545:092.866] SPEAKER_00 I think they're rebased, yeah.
012 - [113.055:113.855] SPEAKER_00 Zyra is doing golem.
013 - [114.075:114.876] SPEAKER_00 Zyra is doing golem.
014 - [122.020:123.141] SPEAKER_00 He was pecking left.
015 - [123.501:124.161] SPEAKER_00 He was leaving.
...
021 - [177.551:178.012] SPEAKER_00 I'll do ignite.
022 - [180.774:181.654] SPEAKER_00 Push base, push base.
023 - [181.975:182.555] SPEAKER_00 Okay.
024 - [185.837:191.862] SPEAKER_00 I'm gonna try to base here.
025 - [191.882:192.862] SPEAKER_00 Okay.
026 - [227.498:228.662] SPEAKER_00 Yes, go push bot now.
\end{verbatim}
We can clearly see that sentence 012 and sentence 013 are exactly the same. But sentence 022 and sentence 024 are not the same but they are talking about the same topic. In fact they are both talking more or less about reseting to base. That is why we used semantic similarity between sentences to tell whether a given sentence has been repetitive. We then have this semantic similarity between 0 and 1. The closest to 1 they are, the closest in terms of meaning these two sentences are. All of the results of the above pieces of converstation can be seen at Figure \ref{fig:duplicate-communication}. In the following result section \ref{subsec:duplicate-com-perf} we interpret this similarity score as a percentage saying \textit{"X sentence is YY\% close to sentence Z"}.\\
Then we have a set of similarity score for each sentence from which we take the maximum. That basically means that \textit{we take the sentence in the previous sentences spoken by our player that is the closest to the current spoken sentence in terms of meaning}.

\subsubsection{Mathematical explanation and solution}
\label{duplicate-com-formal}
In the precedent section we talked about a way of extracting the semantic meaning/similarity of a sentence. There are severall methods to extract such embeddings, like word embedding. Here we take a novel approach of using sentence embedding models \cite{reimers2019sentencebert} that takes into account the context to compute the embedding of a given sentence. With the embedding of the reference sentence, we can then compute the embedding of all the previous sentences that has been spoken by the given player in the last $W$ seconds. By applying this embedding process with sentence transformer \cite{reimers2019sentencebert}, we ensure that the semantic meaning of the sentence is encoded within the dimensions of the embedding vector in a high dimensional space ($\approx 1024$ dimensions). The main goal of projecting our sentences in this high-dimensional space is enable decoding using neural networks or direct Euclidian/linear algebra methods.\\
\\
The most common way to compare if two vectors are close to each other is the cosine similarity. This idea is somewhat the same when we perform retrieval operations in the Open Domain Question Answerign Task \cite{lee2019latent}. By applying this cosine similarity to all the sentences in the last $W$ seconds, we then have a set of $N_{W}$ similarity scores from which we take the maximum similarity score. As its name suggests, the cosine similarity returns the $|\cos(\theta)|$ of the angle $\theta$ between two embedding vectors. The closest to 1 this value is, the closest in terms of meaning these two vectors (i.e. sentences) are.\\
\\
Let $\mathcal{S}_{t}$ the set of all sentences spoken by our player in a window of $W$ seconds before $t$. We have, \\
\begin{equation}
	\mathcal{S}_{t} = \{s_{i}\ |\ i \in [t-W ; t-1]\}
\end{equation}
To compute the similarity between two sentences we compute the cosine similarity of the two sentences embeddings. As previously mentionned, these sentence embeddings are computed via a succession of BERT \cite{devlin2018bert} blocks, forming a sentence transformer \cite{reimers2019sentencebert}. For a sentence $S_t$ spoken at time $t$ we denote $\vec{E_{S_{t}}}$ its corresponding embedding vector. We then have for the cosine similarity of two sentences spoken at $t$ and $t'$ :\\
\begin{equation}
	cosine\_sim(\vec{E_{S_{t}}},\ \vec{E_{S_{t'}}}) = \bigg| \frac{\langle\ \vec{E_{S_{t}}}\ |\ \vec{E_{S_{t'}}}\ \rangle}{\|\vec{E_{S_{t}}}\| \times \|\vec{E_{S_{t'}}}\|} \bigg|
\end{equation}
Where $\langle\ \cdot\ |\ \cdot\ \rangle$ is the dot product of two vectors.\\
We then have the score of all the sentences denoted by it's time $t$ index within 0 and $T_{max}$, where $T_{max}$ is the amount of sentences spoken by our player :
\begin{equation}
	\forall t \in [0, T_{max}],\ Global\_Score = \max_{S_{i} \in \mathcal{S}_{t}}(cosine\_sim(\vec{E_{S_{t}}},\ \vec{E_{S_{i}}}))
\end{equation}
This score basically tells us : \textit{"For the sentence spoken at time t, it has a $s$ score of beeing redundant to the $ith$ sentence preceding it"}\\
You can see in Figure \ref{fig:duplicate-communication} that for each sentences of SPEAKER\_00 we have a score ranging from 0 to 1 that leverage how much this given sentence is being flagged as a repetitive one based on the previous context. And you can see that sentence 024, as mentioned in section \ref{duplicate-com-formal}, has a high similarity score (0.65), which echoes to the semantic meaning of sentence 022.

\subsection{Experiments}
\label{subsec:duplicate-com-perf}
In this section we will take a closer look on how the method detailed in \ref{duplicate-com-explanation} behaves on a given piece of conversation. Please refer to Figure \ref{fig:duplicate-communication-1} alongside reading this section.
\subsubsection{Experimental set-up}
\begin{itemize}
	\item \textbf{$W$} : 15s
	\item \textbf{Embedding model} : "mixedbread-ai/mxbai-embed-large-v1" \cite{emb2024mxbai}
	\item \textbf{Audio length} : 235s
	\item \textbf{Amount of speakers} : 3
\end{itemize}

\subsubsection{Performances/Experiments}
\begin{figure}[h!]
	\centering
	\includegraphics[width=0.85\textwidth]{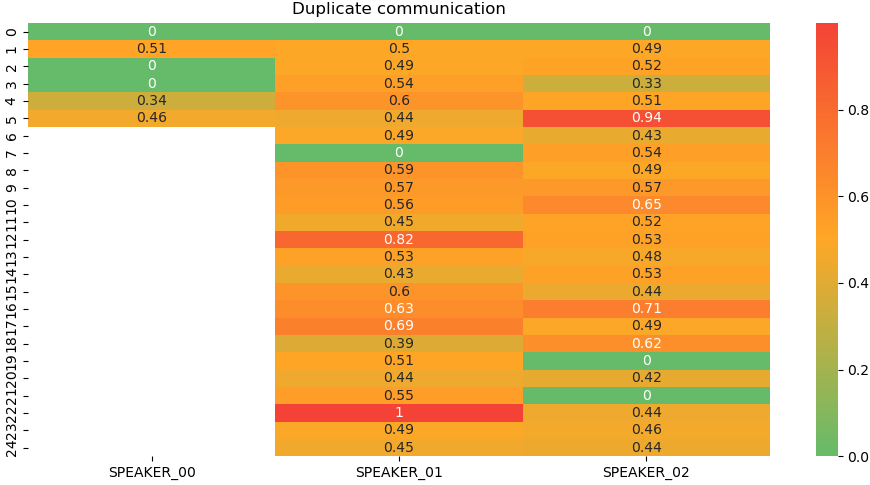}
	\caption{Duplicate communication scores on each sentence for each speaker on another game}
	\label{fig:duplicate-communication-1}
\end{figure}
Here in this section we will take a closer look on another game where we used our tool on. The main objective here is to analyze to determine if the results aligns with the expected outcomes. The transcriptions of Figure \ref{fig:duplicate-communication-1} are listed in Appendix \ref{appendix-communication-logs}. \\
For the purpose of clarity we will only focus on SPEAKER\_01, however feel free to perform the same analysis for SPEAKER\_02 and SPEAKER\_00.\\
On Figure \ref{fig:duplicate-communication-1} most sentences have a score around 0.5, indicating typical communication characteristics. This behavior is widely due to how conversation are done, sentences tend to focus on similar subjects, here being what's happening within the game. However it is noticable that if the score is raising above the 0.6 threshold, the given sentence is somewhat repetitive.\\
Let's take the case of the $12^{th}$ sentence. Here are below the sentences spoken by SPEAKER\_01 15s before sentence 12 :\\
\begin{verbatim}
007 - [69.556:71.557] SPEAKER_01 Okay, I'm moving top side now, okay?
008 - [71.577:73.239] SPEAKER_01 Probably just ward, but he can move top.
009 - [73.259:0074.4] SPEAKER_01 I think you should base and I'll stay.
010 - [074.86:76.141] SPEAKER_01 I can get BF in two waves.
011 - [77.482:78.503] SPEAKER_01 I think we can't, boys.
012 - [79.104:79.864] SPEAKER_01 No, no, no, we can't.
\end{verbatim}

Here it is clearly observable that the $11^{th}$ sentence has a close meaning to the $12^{th}$ sentence, where both sentences have the purpose of holding SPEAKER\_01's teammates back. Also it is not insignificant that both sentence were spoken in a short time frame. The $11^{th}$ sentence was spoken at 77.4s and $12^{th}$ was spoken at 79.1s, which clearly demonstrate the fact that at that time, either SPEAKER\_01 didn't stated clearly his thoughts (see section \ref{sec:ParasiteCommunications} for parasite communication analysis) or SPEAKER\_01's temmates didn't followed his advices.\\

Let's take another example where the score is slightly above 0.6 with sentence 16. Here is the sentences spoken by SPEAKER\_01 15s before sentence 16:\\
\begin{verbatim}
010 - [074.86:76.141] SPEAKER_01 I can get BF in two waves.
011 - [77.482:78.503] SPEAKER_01 I think we can't, boys.
012 - [79.104:79.864] SPEAKER_01 No, no, no, we can't.
013 - [83.027:83.688] SPEAKER_01 Yeah, me too, me too.
014 - [83.728:84.008] SPEAKER_01 No waves.
015 - [87.848:88.868] SPEAKER_01 My mid is going pretty good.
016 - [89.028:92.269] SPEAKER_01 I survived the early game phase, so... What's up, dude?
\end{verbatim}
Here the closest sentence to sentence 16 is sentence 15. In fact they are both more or less talking about the laning phase, but it is clearly not stated and quite blury if both sentences are talking about exactly the same topic. Here the sentence 15 is asserting that SPEAKER\_01 midlane is going well, while the sentence 16 is reasserting this statement. However the sentence 15 could be interpreted differently as it could be stating that the SPEAKER\_01's enemy midlaner is doing well. That is why the similarity score between these two sentences is only at 0.63. \\
This aspect of ambiguity misunderstanding can be due to the fact that the embedding model \cite{emb2024mxbai} was trained on general purpose corpus, and not on League of Legend specific one.

\section{Parasite communications}
As we have seen in section \ref{subsec:duplicate-com-perf} sometimes the way that players express their thoughts might not be clear, causing blurred and unclear communication among players and penalizing in the short term the team's performance. To assess this issue we have tried to use a similar approach based on lexical and semantic similarity \cite{Chen2017odqa}. 
\label{sec:ParasiteCommunications}
\begin{figure}[h!]
	\centering
	\includegraphics[width=0.85\textwidth]{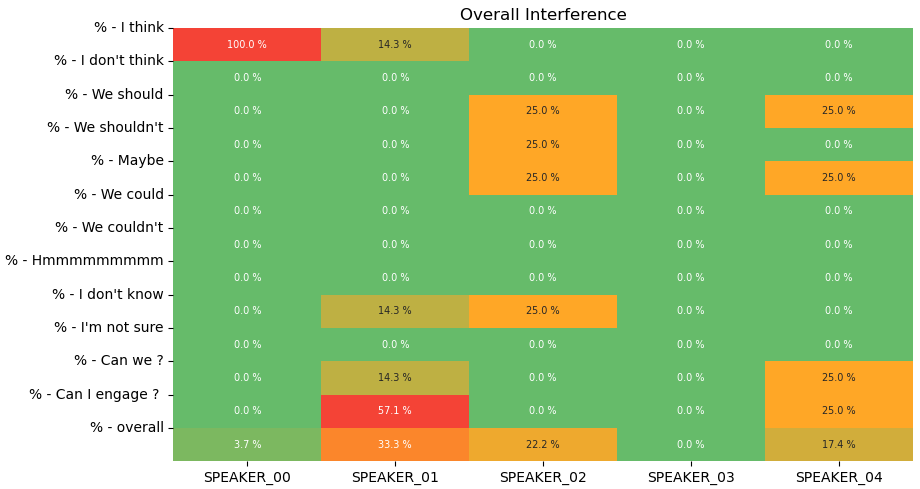}
	\caption{Bottom : Overall interference score of the speaker's communication. Above : interference percentage on each parasite phrasings}
	\label{fig:interference-score}
\end{figure}
\subsection{First approach of the problem}
\subsubsection{Explanation and solutions}
The first aproach of solving this problem was somewhat similar to the one seen in section \ref{sec:DuplicateCommunications}. But to address the inherent uncertainty in such communication, we came up, thanks to the help of professional coaches, with a set of phrasing that would be representative of the unwanted communnication styles (see Appendix \ref{appendix-parasite-phrasings}). For example given the two following sentences:
\begin{verbatim}
	001 - [77.482:78.503] SPEAKER_01 I think we can't, boys
	002 - [77.482:78.503] SPEAKER_01 We can't
\end{verbatim}
These two sentences are inherently saying the same thing, for sentence 001, the phrasing is not appropriate. In fact such phrasing would yield uncertainty in voice communications, hence making communication less effective and directive.\\
This can be verified by comparing the sentences embeddings with the embedding of the phrasing \textit{"I think"} ($P_1$) (extracted from the list of phrasing listed in Appendix \ref{appendix-parasite-phrasings}) with the embedding model of \cite{emb2024mxbai} :\\
\begin{equation}
	\begin{split}
		Score_1 = cosine\_sim(\vec{E_{S_{1}}},\ \vec{E_{P_{1}}}) = 0.5109 \\
		Score_2 = cosine\_sim(\vec{E_{S_{2}}},\ \vec{E_{P_{1}}}) = 0.4027
	\end{split}
\end{equation}
Here we clearly have $Score_1 > Score_2$, which validates the fact that the first sentence is more parasite than the second one. To build such metric we would proceed as following :\\
Let $n_{i}$ the amount of sentences spoken by SPEAKER\_i and $P_j$ the $j^{th}$ parasite phrasing from Appendix \ref{appendix-parasite-phrasings}. We first compute for each sentences spoken by SPEAKER\_i, the similarity score with each parasite phrasings. We have then for each sentence $S_k,\ k \in [0, n_i]$:\\
\begin{equation}
	\mathcal{S}\mathcal{C}_e^{i} = \{cosine\_sim(\vec{E_{S_k}},\ \vec{E_{P_j}})\ |\ j \in [0, n_i]\}
\end{equation}

\begin{figure}[h!]
	\centering
	\includegraphics[width=1\textwidth]{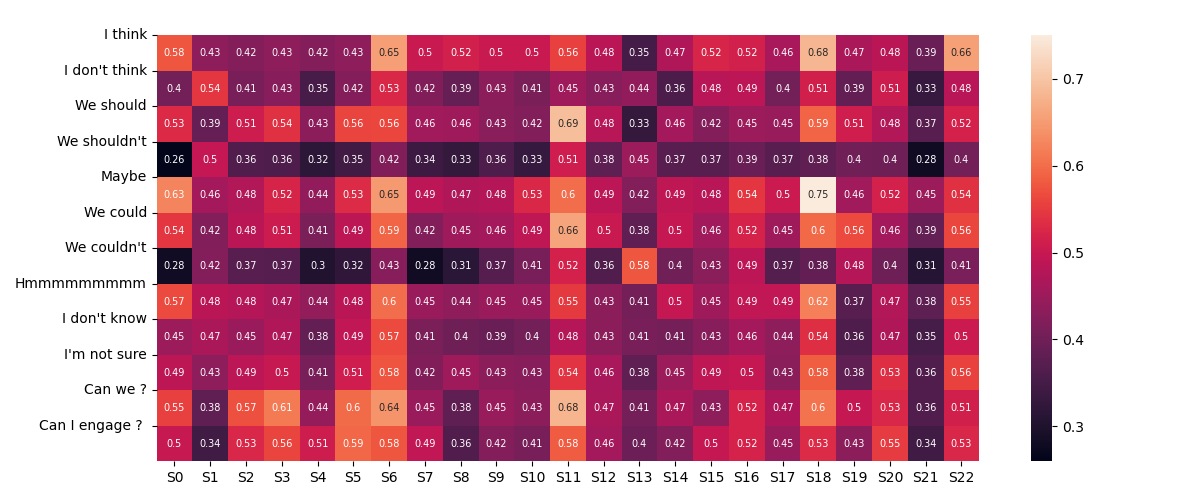}
	\caption{Similarity score of each sentence spoken with eash parasite phrasing}
	\label{fig:interference-heatmap}
\end{figure}

We then take the maximum of these scores and flag it as parasite if it is above 0.6. We denote it as $\mathcal{F}_{S_i} $. Let $ \mathcal{F}$ the set of real number between 0.6 and 1. We have :\\
\begin{equation}
	\begin{split}
		\mathcal{F} = \{ x \in \mathbb{R}\ |\ 0.6 \leq x \leq 1\}\\
		\mathcal{F}_{S_i} = \mathds{1}_{\mathcal{F}}(max(\mathcal{S}\mathcal{C}_e^{i}))
	\end{split}
\end{equation}
Where $\mathds{1}_{\mathcal{F}}(\cdot)$ is the characteristic function of the set $\mathcal{F}$ that yields 1 if the parameter is in the given set and 0 otherwise.\\
With this process, each sentences is flagged at 0 or 1 given if we deem it as parasite. 
Before taking the maximum of $\mathcal{S}_e^{i})$ we have the following heatmap at Figure \ref{fig:interference-heatmap}. On each column of this heatmap we have the similarity score between the given sentence and every parasite phrasing from appendix \ref{appendix-parasite-phrasings}. And it's by analysing the maximum values of each similarity score columns that we flag the sentence as parasite or not.\\

\subsubsection{Limitations}
\begin{figure}[h!]
	\centering
	\includegraphics[width=1\textwidth]{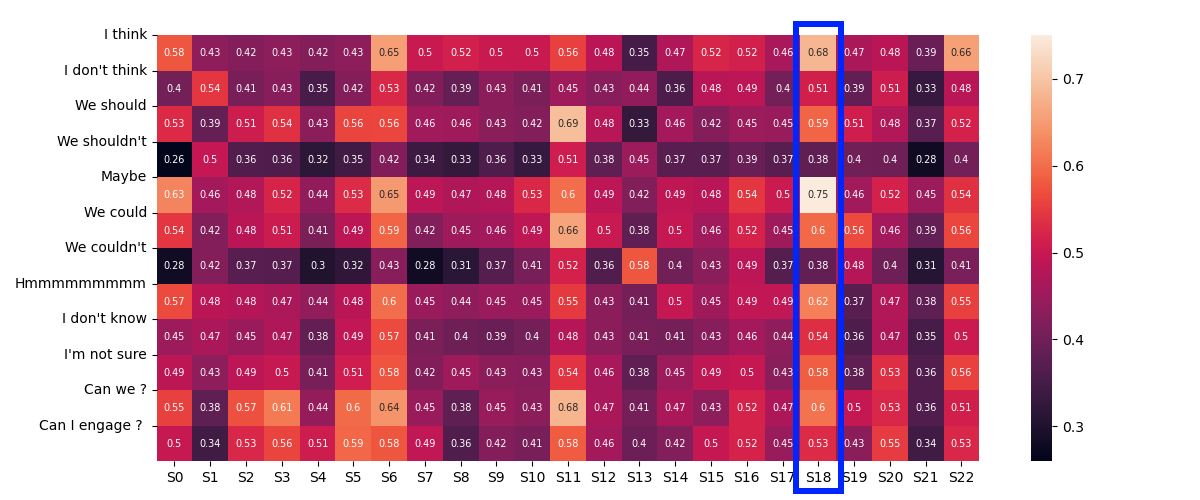}
	\caption{Similarity score of each sentence spoken with eash parasite phrasing}
	\label{fig:interference-heatmap-targetted}
\end{figure}
If we take a close look at sentence 18 (see figure \ref{fig:interference-heatmap-targetted}) it has a quite high similarity score with \textit{Maybe}. Here is what was said at sentence 18 :\\
\begin{verbatim}
	XXX - [YY.YYY:ZZ.ZZZ] SPEAKER_01 Yes.
\end{verbatim}
Given the fact that sentence transformer best perform with full sentences rather than with single words \cite{reimers2019sentencebert} this outlier is predictable because we take a single word sentence as input. The issue here is that the embedding of the word \textit{Yes} varies given the context of the conversation. But given the fact that we don't take the content of the conversation preceding this sentence, it is impossible to encode within the embedding of this sentence the meaning of this sentence in the conversation context. 

\subsection{Refining the first aproach : Embedding refining}
\label{embedding-refining}

\begin{figure}[h!]
	\centering
	\includegraphics[width=0.5\textwidth]{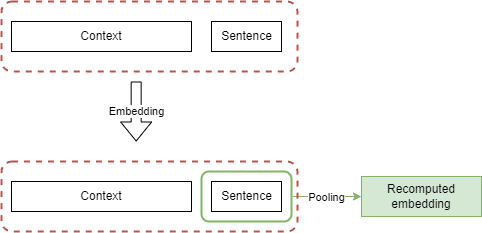}
	\caption{Recomputing the embedding by taking the preceding context and pooling the individual token embeddings of the sentence}
	\label{fig:recomputing-embedding}
\end{figure}

To leverage this issue, we could compute the embedding of the conversation context a fixed time prior to the problematic sentence, then make a pooling operation on the individual token embeddings, that has been recomputed with the context of the converstation, corresponding to the problematic sentence. The overall process is depicted in Figure \ref{fig:recomputing-embedding}

\begin{figure}[h!]
	\centering
	\includegraphics[width=1\textwidth]{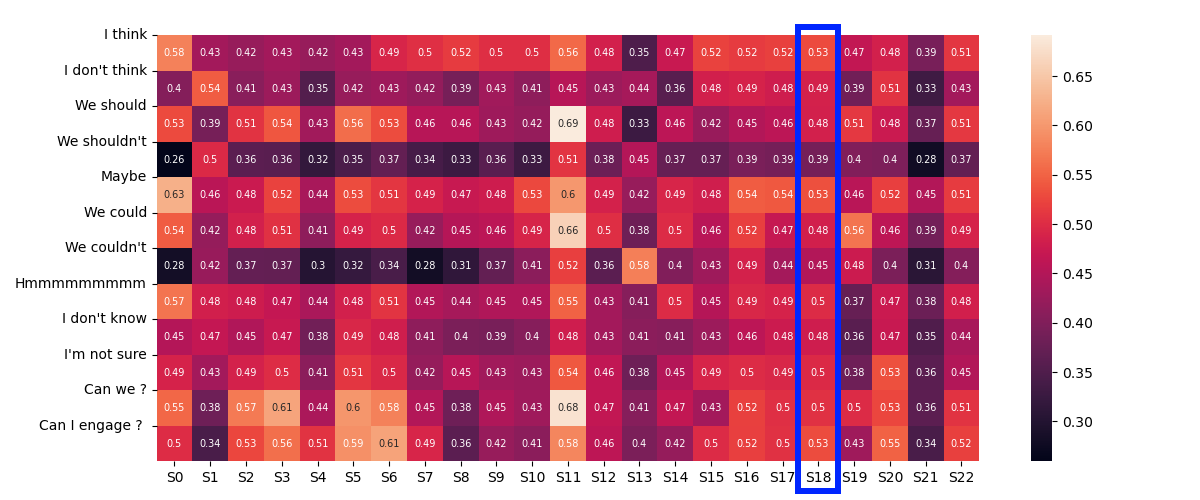}
	\caption{Similarity score sentence embedding, by recomputing embedding of single-word sentences, with eash parasite phrasing}
	\label{fig:interference-heatmap-recomputed-targetted}
\end{figure}

By applying this methodology on the single word sentences we end up with the heatmap shown in Figure \ref{fig:interference-heatmap-recomputed-targetted}

As we can see the recomputed embedding on sentence 018 yields better results when performing sentence similarity compared to the non recomputed one at figure \ref{fig:interference-heatmap-targetted}. In fact the similarity score don't go further than 0.5, which shows that our method managed to distil the context of the sentence 018 answers into the recomputed embedding.

\newpage

\subsubsection{Experiments}
\begin{figure}[h!]
	\centering
	\includegraphics[width=0.8\textwidth]{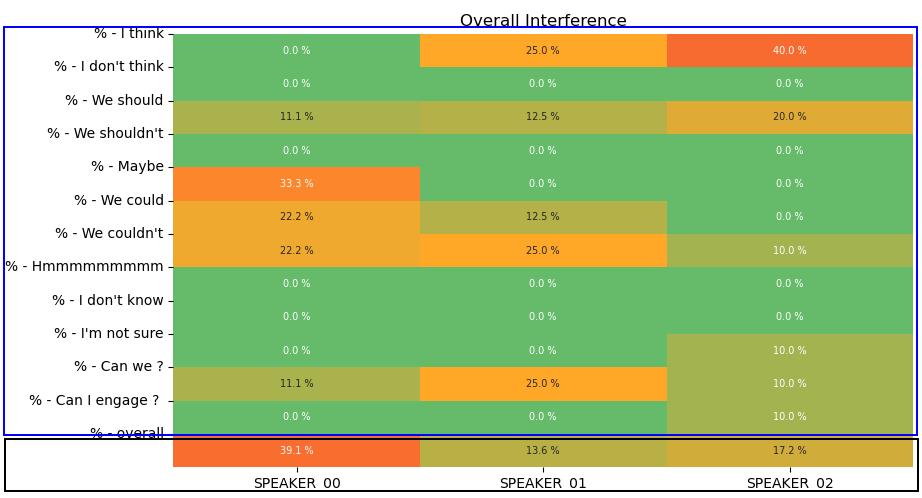}
	\caption{Interference scores of another games}
	\label{fig:interference-score-2}
\end{figure}

For this section we will take a look at the interference of SPEAKER\_00 in the example shown at Figure \ref{fig:interference-score-2}. The sentences spoken as well as the corresponding recomputed interference heatmap are availabale at Appendix \ref{appendix-analysis-materials}. In our example SPEAKER\_00 has spoken \textbf{23 sentences}.\\

On Figure \ref{fig:interference-score-2} we have two parts. The last line (black frame) depicts how many time a given speaker had parasite phrasing in the way that he was speaking. In our case, SPEAKER\_00 was having parasite phrasing 39.1\% of the time. The 12 lines above the bottom one (blue frame) are basically telling us how often, when the speaker is having parasite phrasing, he is refereing to a given emotion/feeling/phrasing. For example, whenever SPEAKER\_00 is having parasite phrasing, 11.1\% of the time he is refering to a phrasing/emotion close to "We should".\\
Let's make the anlysis by hand for sentence 3, 10 and 17 listed bellow :
\begin{verbatim}
003 - [100.811:103.092] SPEAKER_00 If we can... Could we go into Drake?
010 - [261.181:262.082] SPEAKER_00 Wait Lucian, let me pull this.
017 - [355.447:355.707] SPEAKER_00 Okay.
\end{verbatim}
When looking at sentence 3 we can clearly see that SPEAKER\_00 was rather unconfident in his call. And if we look at the interference heatmap at Appendix \ref{appendix-analysis-materials} we can see that this sentence is having the highest score of 0.72 with the phrasing \textit{"Can we ?"}. Which reasonably encompass the feeling of sentence 3. However when lookin at sentence 10, we see that his call is very concise and precise, which is exactly what we want from players when communicating in-game. That's why the scores of sentence 10 don't go higher than 0.56. Lastly to show that the process introduced in section \ref{embedding-refining} works we take a look at the $17^{th}$ sentence, that only contains a single word \textit{"Ok"}. We can see on the interference heatmap that this sentence is not interfering as its scores doesn't go further than 0.56. To put into perspective, when looking at the interference heatmap from appendix \ref{appendix-heatmap-no-recompute} we see that the column coresponding to the $17^{th}$ sentence is having higher interfering scores going up to 0.62. That shows that without recompute this sentence would have been labelled as parasite even though it isn't.
\pagebreak
\section{Performance}
\subsection{Experimental set-up}
\begin{itemize}
	\item \textbf{$W$} : 15s
	\item \textbf{Audio length} : 359s
	\item \textbf{Amount of speakers} : 3
	\item \textbf{Amount of sentences}:  129
\end{itemize}
All the data has been labelized by human and including professional coaches in the process to ensure the quality of labeling.
\subsection{Results}
Here in table \ref{tab:performance}, the results of the experimental set-up above is provided on the duplicate communication model and parasite communication model. For duplicate communication and parasite communicaiton we took a fixed threshold of decision at 0.6.

\begin{table}[h!]
\centering
\caption{Performance Overview and comparison of our model on different sentence transformer architectures}
\label{tab:performance}
\resizebox{\textwidth}{!}{%
\begin{tabular}{@{}lcccccccc@{}}
\toprule
\multirow{2}{*}{\textbf{Model}} & \multicolumn{2}{c}{\textbf{Accuracy}} & \multicolumn{2}{c}{\textbf{Precision}} & \multicolumn{2}{c}{\textbf{Recall}} & \multicolumn{2}{c}{\textbf{F1-Score}} \\ 
\cmidrule(lr){2-3} \cmidrule(lr){4-5} \cmidrule(lr){6-7} \cmidrule(lr){8-9}
 & \textbf{Duplicates} & \textbf{Parasite} & \textbf{Duplicates} & \textbf{Parasite} & \textbf{Duplicates} & \textbf{Parasite} & \textbf{Duplicates} & \textbf{Parasite} \\ 
\midrule
mxbai-embed-large-v1 \cite{emb2024mxbai}& 79.07\% & 83.72\% & 36.84\% & 50.00\% & 82.35\% & 57.14\% & \textbf{50.91\%} & \textbf{53.33\%} \\ 
all-mpnet-base-v2 & \textbf{86.82\%} & \textbf{84.50\%} & \textbf{50.00\%} & \textbf{100.00\%} & 29.41\% & 4.76\% & 37.04\% & 9.09\% \\ 
bge-large-en \cite{bge_embedding}& 29.46\% & 16.28\% & 15.09\% & 16.28\% & \textbf{94.12\%} & \textbf{100.00\%} & 26.02\% & 28.00\% \\
\bottomrule
\end{tabular}%
}
\end{table}

\begin{table}[h!]
\centering
\caption{Embedding Model Rankings from the MTEB Leaderboard \cite{muennighoff2022mteb}}
\label{tab:mteb_leaderboard}
\resizebox{\textwidth}{!}{%
\begin{tabular}{@{}lccccccc@{}}
\toprule
\textbf{Model Name} & \textbf{Rank} & \textbf{Overall} & \textbf{Classification} & \textbf{Clustering} & \textbf{Retrieval} & \textbf{STS} & \textbf{Parameters} \\ 
\midrule
mxbai-embed-large-v1 \cite{emb2024mxbai}& 32 & 64.68\% & 75.64\% & 46.71\% & 60.11\% & 85\% & 335M \\ 
all-mpnet-base-v2 & 115 & 64.23\% & 75.97\% & 46.08\% & 60.03\% & 83.11\% & 335M \\ 
bge-large-en \cite{bge_embedding}& 39 & 57.77\% & 65.03\% & 43.69\% & 59.36\% & 80.28\% & 110M \\   
\bottomrule
\end{tabular}%
}
\end{table}

Table \ref{tab:performance} shows significant performance variation across embedding models, indicating that model selection greatly influences duplicate and parasite communication detection accuracy. In fact these models are trained on general-purpose corpora and not fine-tuned on League of Legends specific corpus. After some surverys with profesionnal coaches, we think it's due to the specific and unique League of Legends' jargon and fast-paced nature of esports communications. This might be the cause of these performance variation, as these models tends to not seize the minute intricacies of the League of Legends vocabulary and way of speaking.\\
We can see that among the tested models, even though other models tends to have better accuracy, precision and recall on each of our tasks, mxbai-embed-large-v1 \cite{emb2024mxbai} showed the most balanced performance, achieving the highest F1-score across tasks, which suggests it best captures the contextual nuances relevant to esports communication.\\
Finally when taking a closer look we see that for model \cite{emb2024mxbai} some performance issues still prevail. Distilling the predictions and comparing them to the ground truth provided us these insights:\\
\begin{itemize}
	\item \textbf{Parasite:} We noticed for False Positive and False Negative that the way that we decide or not to classify a sentence as being parasite with a fixed threshold tends to result in some errors. In a same way sometimes the way that we embed the parasite phrasing/sentiment does not fully represent the full emotional tone of the phrase within the game's context.
	\item \textbf{Duplicates:} For False Positive and False Negative we noticed that the same issue occurs due to the fixed threshold. However we noticed that sometimes, within the same sentence, the speaker is saying the same words several time. A way to leverage this issue would be to use perform an n-gram similarity search within each sentences.
\end{itemize}
All of the potential improvements of our methods will be further described and explained in the conclusion section \ref{conclusion}.

\newpage
\section{Related Works}
The analysis of voice communication, especially within team-oriented environments, has gained our attention as advancements in Natural Language Processing and machine learning provide new tools for interpreting nuanced human interactions. However in this paper, we only took a look at the voice communication aspect of the League of Legends. As explained in the following section \ref{conclusion} we think that correlating these metrics with in-game performance indicators could be relevant to better capture the team performance. That is why we will present some related works that treats the in-game and draft part of the competitive aspect of League of Legends.\\
\\
\textbf{Draft recommendation system:} Some work have been done regarding building a model to recommend champions picks within a draft based on the draft context and the player games history. A dual network approach has been presented by the KAIST \cite{draftRec2022KAIST} paper. One networks aims to reproduce the embedding system of the BERT network \cite{devlin2018bert} by taking from the player's match histroy the champions he played, the roles and some other features in order to generate the embedding of that player profile.\\
Then the other network is a prediction autoregressive transformer-based neural network that recommends the best champions given the draft state and the embedding of our player.\\
\\
\textbf{Predicting match outcome to extract relevant player metrics:} Some other works were related around building prediction model in order to predict the match outcome. A first approach of using machine learning methods and classic architectures with real time statistics (team champion kills, total golds, etc...) was explored by Jailson B. S. Junior et. al. \cite{mlmobapred2023Jailson}. In this article they proceeded of training several machine learning common architectures (Random Forest, Logistic Regression, Naive Bayes, Gradient Boosting, XGBoost, LightGBM, MLP, Bagging and RNN) on these real-time game data to predict the match outcome.\\
Similarly the work from P. Jalovaara \cite{xpetu2024thesis} is using a neural network approach. With some precise tuning on MLP networks and training objectives P. Jalovaara managed to have promising results in predicting match outcome and extending his methods to optimal build path as well. \\
\\
Another notable work is the one from Jiang et. al. \cite{MOBAEmbedding2021Jiang} where they've build a custom embedding system called NICE (Neural Individualized Context-aware Embeddings). This system aims to use contextual information of a given player in a given state of the game to predict the match outcome. In order to do this they generate the embeddings from a set of features of these sets : \\
\begin{equation}
	user \times global\_context \times individual\_contexts
\end{equation}
All this performed with the help of the Non-Negative Tensor Factorization \cite{nntf2009Kolda} method.

\newpage

\section{Conclusion and discussion / improvements}
\label{conclusion}
In this paper, we presented an approach to analyze voice communications in esports, specifically focusing on detecting duplicate and "parasite" communication in League of Legends. Through the use of semantic similarity measures and NLP embedding techniques, we developed metrics to assess communication quality and its potential impact on team performance. While the results offer promising results some dark parts still remains to be assessed that might be subject of future research.

\subsection{Improvements}
\textbf{Specialized Embedding Model:} One of the limitation of our current approach is that it relies on embeddings from models trained on general-purpose corpora. In fact in this paper we used such general-purpose embedding model from \cite{emb2024mxbai, bge_embedding}. Future work could focus on developing an embedding model trained specifically on League of Legends or other esports related datasets. This would likely capture the unique linguistic patterns, terminology, and contextual cues in esports, potentially improving the model’s ability to identify nuanced, context-sensitive communication.\\
\\
\textbf{Correlation with In-Game Performance:} Currently, the relationship between communication quality and team performance remains unmentioned in our analysis. Future research could try to incorporate in-game performance metrics with communication metrics. Hence trying to build multimodal metrics that could encapsulate better player and team performance. This approach could reveal more direct associations between communication effectiveness and game outcomes, providing empirical validation on the impact of voice communication on team success.\\
\\
\textbf{Raw Audio Analysis:} At first we focused on applying NLP techniques on transcribed text as it easier than treating raw audio file. That is why extending this methodology to include raw audio data could provide a richer understanding of communication dynamics. By leveraging audio features such as tone, pitch, and volume, we might capture additional elements of speaker intent and sentiment that text alone cannot convey. A way of such integreation would be to use whisper's encoder \cite{openai2022whisper} to generate audio embedding of each pieces of speech. Integrating these audio features could create a multimodal analysis model, enhancing the detection of key communication traits. However the problem of using general-purpose audio model still remains when applying it on specialized League of Legend data.\\
\\
\textbf{Improved Detection of Parasite Sentences:} Our current decision function for flagging "parasite" sentences is binary and based on a fixed similarity threshold. In our work we chose 0.6 as it was one of the most reliable threshold. To refine this, we could implement a smoother decision function that adjusts the threshold based on the context and speaker. By using a continuous rather than binary function, the model might better discriminate between minor conversational nuances and genuinely disruptive communication patterns, enhancing its sensitivity to context. This way we could try to reproduce the linear and smooth activation function after a linear layer in an MLP network, but adapted as a decision function for "parasite" activation on a given piece of speech.\\
\\
\textbf{Enhanced parasite phrasing/sentiment encoding:} The current model considers only short, isolated pieces of sentences/phrasing to detect a parasitic speech. Future work could involve a more sophisticated parasite phrasing/sentiment encoding mechanism that captures the broader emotional tone of phrases within the game’s context. For instance, pooling embeddings of multiple related words or phrases could yield more accurate  parasite phrasing/sentiment vectors, improving the model's ability to assess both positive and negative influences of communication on team dynamics.\\
\\
\textbf{New metric sentence relevance:} Currently our parasite sentence detection only takes into account sentences that reflects hesitation and/or uncertainty in the way it is spoken. However it does not takes into account the fact that players might talk about irrelevant topics during the game, hense making the team lose focus on the said game. One approach would be to also use sentence similarity but this time comparing the vector embedding of the current sentence with the embedding of the conversation 15s prior to the said sentence. By doing this we could have a metric that could ensure how often each player are talking about topics that are not directly related to the game's state.\\
\\
In summary, while our study provides a foundational framework for analyzing voice communication in esports, these potential improvements represent valuable opportunities for refinement. Continued research in these areas could contribute to a more holistic and nuanced understanding of how communication impacts team performance in competitive gaming.

\subsection{Practical Contributions}
Beyond the theoretical insights provided, this work offers tangible applications that can aid coaches in optimizing team performance through improved communication analysis. By applying the voice-communication metrics developed here, coaches can gain a clearer understanding of each player’s communication profile, identifying tendencies like repetitive or unclear communication that might by a liability for the overall team cohesion. This profile-based insight can help coaches adapt their feedback to suit each player’s unique communication style, fostering better synergy in team interactions.\\
\\
Furthermore, this analysis provides a comprehensive overview of team communication as a whole, allowing coaches to assess how effectively the team communicates in tense situations. By visualizing patterns of redundant and parasite communications, coaches can identify specific areas where the team excels or struggles, giving them a baseline measure of team coordination that can be optimized over time.\\
\\
Finally, the insights from this study can serve as a valuable resource for planning future training sessions. Coaches can target identified weaknesses in communication, structuring training exercises to address specific issues such as reducing redundant calls or encouraging more direct communication during critical in-game moments. This approach not only enhances communication quality but also ensures that each practice session is strategically aligned with the team’s communication needs.
\newpage
\bibliographystyle{plain}  
\bibliography{references}  

\newpage
\section{Appendix}
\subsection{Communication Logs}
\label{appendix-communication-logs}
\begin{verbatim}
000 - [00.942:02.163] SPEAKER_01 You need to push out, XXXX.
001 - [02.884:03.424] SPEAKER_01 Oh, okay.
002 - [03.684:05.826] SPEAKER_01 They do my Gromp.
003 - [17.837:18.618] SPEAKER_01 I'll face Zyra, Leona.
004 - [19.078:22.741] SPEAKER_01 I'm going.
005 - [23.622:23.882] SPEAKER_01 Can you?
006 - [30.828:32.469] SPEAKER_01 Okay, not bad, got a kill WP.
007 - [69.556:71.557] SPEAKER_01 Okay, I'm moving top side now, okay?
008 - [71.577:73.239] SPEAKER_01 Probably just ward, but he can move top.
009 - [73.259:0074.4] SPEAKER_01 I think you should base and I'll stay.
010 - [074.86:76.141] SPEAKER_01 I can get BF in two waves.
011 - [77.482:78.503] SPEAKER_01 I think we can't, boys.
012 - [79.104:79.864] SPEAKER_01 No, no, no, we can't.
013 - [83.027:83.688] SPEAKER_01 Yeah, me too, me too.
014 - [83.728:84.008] SPEAKER_01 No waves.
015 - [87.848:88.868] SPEAKER_01 My mid is going pretty good.
016 - [89.028:92.269] SPEAKER_01 I survived the early game phase, so... What's up, dude?
017 - [094.27:095.07] SPEAKER_01 Bot is going pretty good.
018 - [105.152:106.713] SPEAKER_01 Uhh... I have a tree in my back.
019 - [107.513:107.693] SPEAKER_01 Okay.
020 - [107.713:112.554] SPEAKER_01 I mean, we don't have imp there, so... Everyone moving, yeah.
021 - [119.443:120.163] SPEAKER_01 I will not push this.
022 - [120.784:121.044] SPEAKER_01 Okay.
023 - [124.826:130.829] SPEAKER_01 He wants to W from here, but... Any flash, XXXX?
024 - [138.813:140.174] SPEAKER_01 I'm basically late, but I'm really strong.
\end{verbatim}

\subsection{Parasite Phrasings}
\label{appendix-parasite-phrasings}
\begin{itemize}
	\item I think
	\item I don't think
	\item We should
	\item We shouldn't
	\item Maybe
	\item We could
	\item We couldn't
	\item Hmmmmmmmmm
	\item I don't know
	\item I'm not sure
	\item Can we ?
	\item Can I engage ?
\end{itemize}

\subsection{Materials for analysis}
\label{appendix-analysis-materials}
\subsubsection{Logs}
\begin{verbatim}
000 - [50.899:55.501] SPEAKER_00 Okay guys, whenever someone flash, say it, I'm pinging it, 
because I got it, so it's easy for me.
001 - [81.726:82.547] SPEAKER_00 I mean, I can't.
002 - [99.511:100.551] SPEAKER_00 No, they just go Void.
003 - [100.811:103.092] SPEAKER_00 If we can... Could we go into Drake?
004 - [103.432:104.012] SPEAKER_00 I miss or no?
005 - [180.564:181.205] SPEAKER_00 I'm dying, I'm dying.
006 - [236.328:236.768] SPEAKER_00 Nice try.
007 - [238.028:241.67] SPEAKER_00 Only XXXXX flashed?
008 - [256.878:258.659] SPEAKER_00 I think get some items and then we can fight again.
009 - [258.799:260.18] SPEAKER_00 Now, about fight maybe.
010 - [261.181:262.082] SPEAKER_00 Wait Lucian, let me pull this.
011 - [262.142:263.162] SPEAKER_00 I get, I get one more.
012 - [335.402:336.643] SPEAKER_00 When Nami is there, we can TP.
013 - [337.624:338.204] SPEAKER_00 If they hit turret.
014 - [347.841:348.621] SPEAKER_00 They can swap him maybe?
015 - [348.961:349.622] SPEAKER_00 Can they swap him?
016 - [350.323:351.143] SPEAKER_00 Yeah, they can.
017 - [355.447:355.707] SPEAKER_00 Okay.
018 - [356.748:357.228] SPEAKER_00 I'll stop him.
019 - [357.368:357.789] SPEAKER_00 I'll stop him.
020 - [357.829:360.271] SPEAKER_00 He's here.
021 - [361.732:362.633] SPEAKER_00 He's still not there, okay?
022 - [362.653:363.434] SPEAKER_00 If he's there, he will TP.
\end{verbatim}
\subsubsection{Interference Heatmap Recomputed}
The max interfering scores that are higher than 0.6 are highlighted in cyan for clarity purposes.
\begin{figure}[h!]
	\centering
	\includegraphics[width=1\textwidth]{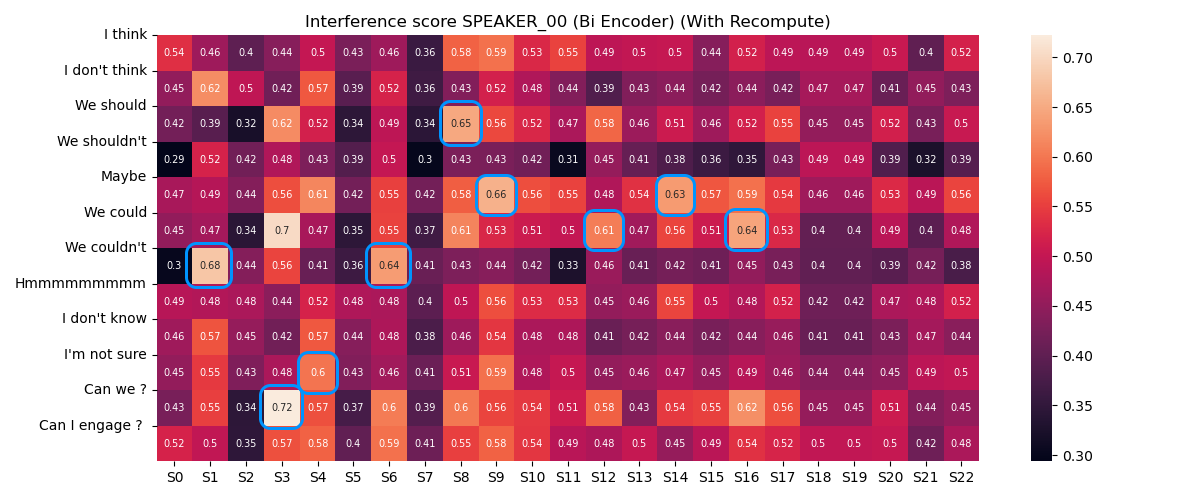}
	\caption{Interference Heatmap of SPEAKER\_00}
\end{figure}

\newpage
\subsection{Interference Heatmap Not Recomputed}
\label{appendix-heatmap-no-recompute}
This heatmap is computed from the same audio sample of above heatmap, but without the embedding refinment

\begin{figure}[h!]
	\centering
	\includegraphics[width=1\textwidth]{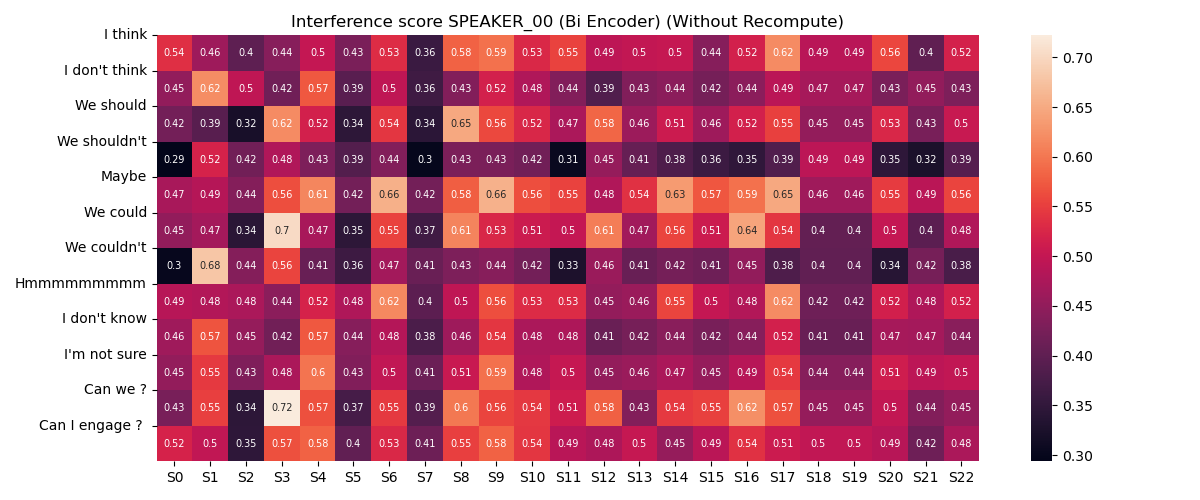}
	\caption{Interference Heatmap of SPEAKER\_00 without embedding refinment}
\end{figure}

\end{document}